# Privacy in Location Based Services: Primitives toward the Solution

Abedelaziz Mohaisen and Dowon Hong
Cryptography Research Team
Electronics and Telecommunication Research Institute
Daejeon 305-700, Korea
{a.mohaisen,dwhong}@etri.re.kr

DaeHun Nyang
Information Security Research Lab
Inha University
Incheon 402-751, Korea
nyang@inha.ac.kr

**Abstract**

*Location based services (LBS) are one of the most promising and innovative directions of convergence technologies resulting of emergence of several fields including database systems, mobile communication, Internet technology, and positioning systems. Although being initiated as early as middle of 1990's, it is only recently that the LBS received a systematic profound research interest due to its commercial and technological impact. As the LBS is related to the user's location which can be used to trace the user's activities, a strong privacy concern has been raised. To preserve the user's location, several intelligent works have been introduced though many challenges are still awaiting solutions. This paper introduces a survey on LBS systems considering both localization technologies, model and architectures guaranteeing privacy. We also overview cryptographic primitive to possibly use in preserving LBS's privacy followed by fruitful research directions basically concerned with the privacy issue.*

## 1. Introduction

The location based services (LBS) are convergence of different technologies resulting of the recent development of the mobile communication and computing, Internet technology, geographical information systems (GIS) [47], spatial database systems, and others. Though the LBS as a technological direction was initially innovated at the end of the last century, it is only a few years ago that it received a strong interest of research due to its applications that promise a huge commercial impact in the major technology consuming and producing regions in the world. For example, according to ZDnet [51], the market size of LBS in Europe will grow to 622 million € in 2010 with an annual increment of 34% from 2006 and the users to grow to 315 million from the current 12 million users all around the world for the same period. In Asia, according to INDOORLBS [22], the market size as of 2006 was 291.7 million US dollars and will be growing at the end of 2009 to count for 447 millions with an annual growth of 15.3% where Japan and Korea share 92% of the market size due to their assisting IT infrastructure. The applications of LBS systems include different services with ranging profit fields. To mention some example, recent LBS applications include location-based traffic report, store finder, and advertisement, among others. In the following we discuss the main scenarios and architectural view of these applications.

The basic scenario of the LBS assumes the existence of three distinct entities which are the LBS server or LBS provider, the mobile operator, and the mobile user. The flow of the LBS, as shown in Figure 1, goes as follow: first the mobile user request the LBS server to provide him or her the available specific set of services within his/her location's area and directions to get into them. Upon that, the LBS server contact the concerned mobile provider to retrieve the user's location. The mobile provider, using of the location technologies that will mentioned later, determines the user's location and send it back to LBS server which make a query listing the available service within the user's location and send it back to the user. More details and the flow of other technique is shown in section 2.

Obviously, to be able to provide the specific set of services for the user within a reasonably accessible range from the user's current location, the user's location is needed by the LBS server. However, providing the exact location would breach the users privacy. Accordingly, several mechanisms are systematically studied in order to provide both service and privacy to the user in the LBS system. These mechanisms are classified according to the rule of the user and underlying technique for hiding the users location. Architecturally, the mechanisms guaranteeing privacy are classified according to the users role into: non-cooperative mechanisms, centralized trust third party mechanisms, and peer-to-peer cooperative mechanisms. According to underlaying technique for hiding the user's location, these mechanisms are classified into cryptographic and non-cryptographic techniques. More details on these techniques will be shown later in this article.

This paper introduces a survey on the LBS systems demonstrating the used technologies and techniques and detailing cryptographic primitives to be used in order to provide a favorable privacy. The rest of this paper is organized as follows: 2 introduces the localization techniques used in LBS systems, section 3 introduces the issues of the privacy in LBS systems considering architectures and R&D projects, section 4 surveys main cryptographic primitives, section 5 introduces directions for further research and section 6 draws concluding remarks.

## 2. Localization techniques

Localization techniques are classified in a high level and low level categories. This sections review the classification

### 2.1. High level classification

High level classification considers the rule of the user where the techniques are into three parts which care networks-based, handset-based and hybrid locations. The **network-based localization** uses existing infrastructure of the mobile operator (i.e., provider) in order to measure the user's current location. The techniques used for measuring the location range from the very accurate methods (e.g., *triangulation*) to the least accurate methods (e.g., *cell-ID*). The accuracy of the location determined using these methods is mainly based on the concentration of the base stations within the network. Therefore, high accuracy is achievable in the urban areas while less accuracy is possible in the sub-urbans. The main challenge of this technique, however, is that the LBS provider needs to work closely with the mobile provider to install hardware and software within the operators infrastructure. While on the other side it requires a *legislation infrastructure* in order to compel the cooperation of the service provider and *safeguard* the privacy of the different users. Examples of such frameworks include **E911** [17] in the US and **E112** [14] in Europe. An illustration of the LBS flow using this technique is shown in Figure 1.

The **Handset-based localization** uses the handset itself in order to determine the user's location. It requires installation of software or hardware pieces on the handset in order to make it able to compute the mobile user's location. The mainly challenge when using this technique is the requirement of an active cooperation of the mobile subscriber as well as the software that must be able to handle different OS setting and types. Currently, only *smart phones* are able to run that software. The used techniques for handset-based localization also range in accuracy and include signal strength computation of the home and neighboring cells, latitude and longitude determination (i.e., when using the GPS), and others. An illustration of the LBS flow using this technique is shown in Figure 2.

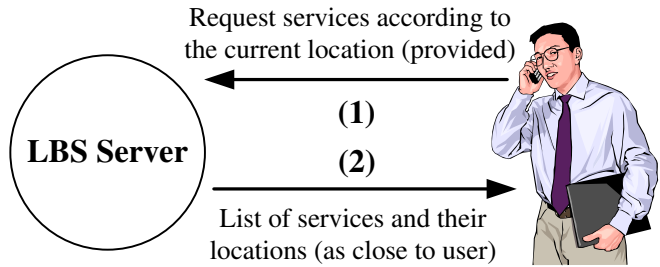

**Figure 2. Flow of** LBS **using handset-based localization model**

Finally, the **hybrid-based localization**, uses both of the above techniques to determine the mobile user's location. While having the advantage of both techniques and providing the highest accuracy, it also inherits the shortcomings of both schemes. An example of the hybrid-based localization is the A-GPS (assisted GPS).

### 2.2. Low level description

A lower level classification of these techniques is done according to the used localization technology rather than the initiator or performer of localization. This classification is shown in Figure 3 and detailed as follows (further details on these techniques and others are detailed in [48]):

- Global Positioning System (GPS): uses a global navigation satellite system (GNSS) utilizing more than 24 medium Earth orbit satellites that transmit microwaves enabling GPS-receivers to determine their location, speed, direction and times. Other similar systems include the Russian GLONASS [41], the EU's Galileo [13], and the Chinese COMPASS navigation system (extension BEIDOU [42]).
- Assisted GPS (A-GPS): uses both the network- and handset-based techniques to determine the user's location. This techniques is mainly used in urban areas.
- Angle of Arrival (AOA): is based on the angle at which device's transmitted signal arrive at the based station.
- Time difference of arrival (TDOA): works on determining time difference and therefore the distance from each base station to the mobile phone.
- Time of Arrival (ToA): similar to the TDOA but differs in that the base station uses the absolute time rather than the difference to compute the distance.
- Observed Time Difference (OTD): is based on the time difference between sending and receiving user's signal and translating the difference into distance.

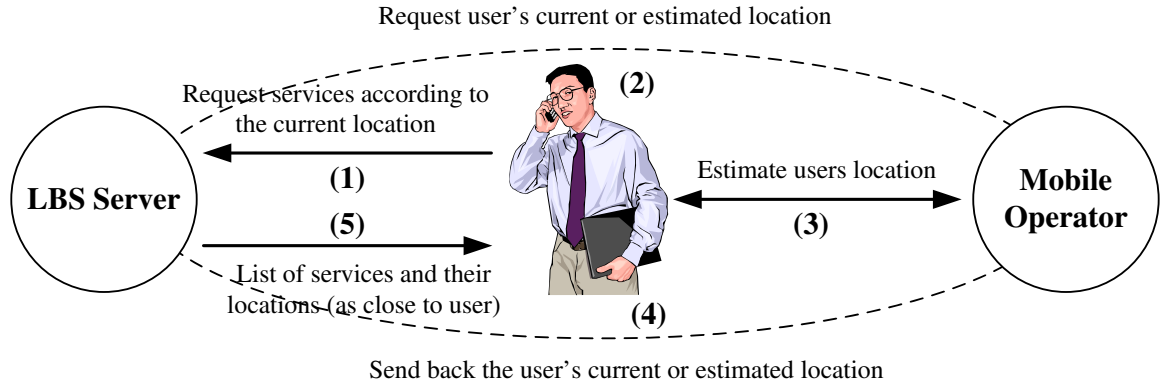

**Figure 1. Flow of LBS using network-based localization model**

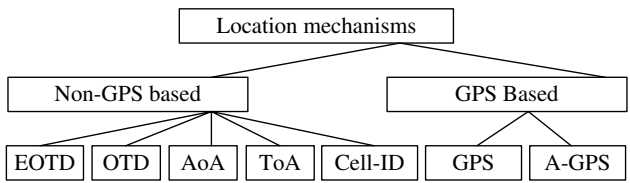

**Figure 3. Location measuring techniques**

- Enhanced OTD - (EOTD): is similar to the TDOA where the time is measured by mobile device but not the base station. Accuracy from 50 to 200 meters.
- Cell-ID: Measures the location of the user according to its nearest mobile cell; cheapest and the least accurate. Currently used by most LBS providers in Europe.
- Enhanced Cell-ID: with this technique, one can obtain same accuracy as of the Cell-ID but in *rural* areas.

## 3. The Privacy Issue

The need for privacy in the LBS systems is very obvious and critical. Since the location information of the user can be used to trace the user's activities and pattern of life, it is necessary to keep such information private from possible adversary. In order to legislate this, several laws, acts, and directions have been made. All of these laws are not limited to the LBS but any system that carries private data. Examples of these acts include (1) Health Insurance Portability and Accountability Act (HIPAA) in the United States of America [45], (2) Canadian Personal Information Protection Act (PIPEDA) in Canada [9], (3) Directive 95/46/EC on the protection of personal data in the European Union [15], and (4) ISO/TC 215, a standardization initiative for general regulations defining private/personal data and usages.

### 3.1. Architectures Achieving Privacy

There are three different architectures for achieving the privacy in the LBS. These architectures are classified according to the rule of user and existence of third party as follows:

1. **Non-cooperative model:** In the model, the user uses his own capability and knowledge to hide his location using one of the hiding techniques. These hiding techniques include Pseudonimity, false dummies and landmark objects. This model is also known for its simplicity in design and vulnerability to several attacks.
2. **Centralized trust third party (TTP) model :** This model relies on using a trust third party for performing the heavy work in determining the proper technique for anonymizing the location of the user, requesting the service with the anonymized location and returning the result to the user. Though being the most accurate and achieving the highest privacy level, this model suffers from being very sophisticated in its design where the TTP represents a challenging bottleneck.
3. **P2P cooperative model:** In this model, several users tries to cooperate and compute their location or hide their location information in a distributed manner in order to achieve a proper privacy. In this model, the cryptographic primitives (such like OT, adaptive OT, proxy OT, and others) play an important rule in secure computation. Also certificates can be used for approving the trustworthiness of the users.

### 3.2. Related R&D Projects

Several research and developments projects have been performed or currently running in order to protect the privacy of individuals users in different networking settings including the LBS systems. This section summarizes some of these efforts.

1. **Privacy and Identity Management for Europe (Prime) [38]:** This project was supported by the European commission under the FC6 support grants and ran from March 2004 to February 2008. Participants to this projects included 18 R&D institutes and research centers which mainly from or operating in Europe. Topics covered in this project included access and policy control, contents and semantics, *location and communication*, *cryptography in LBS* and vehicular networks, trust and reputation management, protection and anonymity, and anomaly credentials. Results in this work included several innovative solutions for some of the existing problems in the above issues, several prototypes, and initiative for several standardizations efforts.
2. **PrimeLife:[39]** is an extension of the above project and based on a joint fund from FP7 and ICT running from March 2008 for 36 months. Participants in this project include 14 research and developments institutes from both the industry and academia. Participants are mainly from Europe and the main topic of research concerned with protection and enhancing privacy of users in next generation Internet environment services including WEB 2.0, mobile Internet, etc. Though not directly related to the LBS, the project dealing with the mobility as an important theme of the next generation Internet may relate to theme of the LBS in a way or another.
3. **Privacy in Ambient Word (PAW) [36]:** This project is made to answer several questions such like: Is it in theory possible to protect mobile software against privacy and security attacks while these programs are executed at untrustworthy hosts operated by the user or operated by an unknown user. Also, are conventional cryptographic algorithms applicable in mobile environments. Several works have introduced in this directions and mainly focused on theoretical and cryptographic bases. The project ran from 2004 to 2008 (completed).
4. **FIDIS network of excellence (FIDIS) [18]:** FIDIS (abbreviation for Future of IDentity in the Information Society) is a network 24 European industrial and academic institute that perform research towards tackling the problems related to the ID management with concentration on the following sub-areas: identity management (foundations), interoperability of identity and identity management systems (IDMS), profiling, forensic implications of IDMS, privacy and legal-social issues of identity, mobility and identity, ID-theft, privacy and security, among others. Though this is not directly related to the LBS and its privacy, the identity (as a threat of privacy) is a general concept which apply to several settings include that of LBS.

## 4. Cryptographic Primitives

The cryptographic direction as a solution for guaranteeing privacy in the LBS systems has not been used yet. Though, a great deal of work have been before on the cryptographic primitives in other networking and technological settings that are awaiting a practical use in LBS. In the following we detail a set of these primitives that have a strong and promising relationship with the LBS.

### 4.1. Blind signature

The blind signature is introduced firstly by David Chaum [10] and it works as follows: Firstly, the owner of the message produces the product of the message and a blinding factor $r^e$ as $m^{'} = mr^e \mod N$ where $m^{'}$, the blinded message, does not reveal any information about the original message $m$. Then, the signing authority signs, the message as follows to reveal $s$ as follows $s^{'} = (m^{'})^d \mod N$. After that, $s^{'}$ is returned to the author of the message who can remove the blinding factor performing the following $s = s^{'} * r^{-1} \mod N$. Obviously this works because the property of the RSA keys satisfying $r^{ed} \equiv r \mod N$. I.e., $s \equiv s^{'} * r^{-1} \equiv (m^{'})^d r^{-1} \equiv m^d r^{ed} r^{-1} \equiv m^d r r^{-1} \equiv m^d \mod N$, where the final result is the typical RSA signature.

The applicability of the blind signature to the privacy is very direct and essential. Assuming cooperative scenario for introducing privacy, any device in the LBS system using the blind signature technique still sign a message from any LBS device without having access to the contents of the message itself (guaranteeing the privacy of its owner). Extensive work has been performed on blind signatures on the context of e-voting and e-cash among others [46, 11, 49, 29].

### 4.2. Oblivious transfer (OT)

The oblivious transfer (OT) protocol is a method used for secure computing in which the sender sends some information to the receiver but remains oblivious to what the receiver has received. [34]. Broadly, the OT protocols are divided into two main protocol categories which are $1 - 2$ OT and $1 - n$ OT (read as 1 out of 2 and 1 out of $n$ respectively). The $1-2$ OT, firstly discussed by Even et al. in [16] and based on RSA [40] assumes a sender which has $m_0$ and $m_1$ as messages and receiver that has a bit $b$. The sender wants to make sure that the receiver receives one message only among his messages ($m_b$) and the sender wants to receive $m_b$ but without making the sender knows the bit $b$ itself. Technically, it works as follows:

1. The sender generates RSA keys, including the modulus $n$, the public exponent $e$, and the private expo-

nent $d$, and picks two random messages $r_0$ and $r_1$, and sends $n, e, r_0$, and $r_1$ to the receiver.
2. The receiver picks a random message $k$, encrypts $k$, and adds $r_b$ to the encryption of $k$, modulo $n$ (i.e., $E(k) + r_b \mod n$), and sends the result $q$ to the sender.
3. The sender computes $k_0$ to be the decryption of $q - r_0$ and similarly $k_1$ to be the decryption of $q - r_1$, and sends $m_0 + k_0$ and $m_1 + k_1$ to the receiver.
4. The receiver knows $k_b$ and subtracts this from the corresponding part of the sender's message to obtain $m_b$.

The $1-n$ OT is a generalization of the the $1-n$ with the assumption that the sender has $n$ messages sorted according to an index $i$ [44]. In that case, the sender want the receiver to know a single message among the $n$ message while the receiver is interested in not revealing the index of the message $i$. This is specially important when the scenario is applied on data retrieval that preserve the privacy of the data. Other variants of oblivious transfer schemes which applies for the distributed systems in general and LBS systems in specific include the following:

- Adaptive OT [32]: This technique differs from the technique mentioned above in that it can be applied adaptively and allow multiple execution without revealing any information exchange.
- Dynamic OT [24]: The dynamic OT runs over a dynamic database that shrinks and grow according to the deletion and addition of data from or in it respectively.
- Proxy OT [50]: The proxy OT is very suitable for devices with limited computation capabilities. That is, the OT procedure which is known to be computationally expensive is applied on another device with a strong computational capability and the final result is returned to the mobile device in order to be used. This is critical in our work as most of devices used in the LBS system are with a limited computational capabilities.

The OT is the building block of the secure multi-party computation and used heavily in other directions for privacy preserving technologies in other areas including the privacy preserving data mining. Scenarios for applications should exist for applying the OT and SMC both for the privacy preserving LBS as well. For more details on the distributed OT and its security issues, please refer to detailed works such like [5], [33], and [12] (the last work details the $k$-out-of-$n$ OT) .

### 4.3. Broadcast Authentication

The broadcast authentication [37] has received a great deal of efforts in the cryptographic society along the known broadcast encryption technique. These efforts are motivated by the promising applications and great commercial impact of this technology. On the devices with limited capabilities, the broadcast authentication have been studied in the context of sensor network and basically based on the authenticating previous messages by delaying the release of their encryption keys as in TESLA and its variants (cf., [25], [26], [27], among others).

### 4.4. Aggregate signatures

The aggregate signature [20] is a digital signature that supports aggregation. That is, given $n$ signatures on $n$ distinct messages from $n$ distinct users, it is possible to aggregate all these signatures into a single short signature. This single signature will convince the verifier that the $n$ users did indeed sign the $n$ original messages.

The bilinear aggregate signature is an example of the aggregate signature and is based on the short signature which works as follows:

- Key Generation: In this phase, public and private keys are generated for the specific users: (a) $x \leftarrow Z_p$ (private key), and (b) $v \leftarrow g^x$ (public key).
- Signing: given $x$, message $M \in \{0,1\}^*$, compute $h \leftarrow H(M)$ where $h \in G$ and $\sigma \leftarrow h^x$, the signature $\sigma \in G$.
- Verify: given $v, M, \sigma$, compute $h \leftarrow H(M)$ and verify $(g, v, h, \sigma$ are valid DH tuples. That is, $e(h, v) = e(\sigma, g)$. For the left side, $e(h, v) = e(h, g^x) = e(h, g)^x$. For the right side, $e(\sigma, g) = e(h^x, g) = e(h, g)^x$ which holds.

The aggregate signature considers the above short signature, $n$ different users with $n$ different messages, and two additional steps which are signature aggregation and aggregate verification. In the BLS aggregate signature for example [8], after performing the three steps above on each user's message, the user performs the following

- Aggregation: given the signatures $\sigma_1, \ldots, \sigma_n$ correspond to the messages $M_1, \ldots, M_n$, compute the signature $\sigma \leftarrow \prod_{i=1}^n \sigma_i, 1 \le i \le n$.
- Aggregate verification: given $n$ different messages $M_1, \ldots M_n$ and public keys $v_1, \ldots, v_n$ compute $h_i \leftarrow H(v_i, M_i)$ and then verify the aggregate signature by accepting if $e(\sigma, g) = \prod_{i=1}^n e(h_i, v_i)$. The verification of this is very straightforward. Take the right side $e(\sigma, g) = e(\prod_{i=1}^n h_i^{x_i}, g) = \prod_{i=1}^n e(h_i^{x_i}, g) = \prod_{i=1}^n e(h_i, g)^{x_i} = \prod_{i=1}^n e(h_i, g^{x_i}) = \prod_{i=1}^n e(h_i, v_i)$ which holds.

For an interesting survey on aggregation signature techniques and schemes with applications, see Boneh's et al. work in [7]. Examples of aggregate signatures that fit to the LBS include works in [6, 31, 28, 4, 30], among others.

## 4.5. Aggregate MAC

The aggregate message authentication code (aggregate MAC) [23] is a special kind of message authentication code that support MAC aggregation. The message authentication code, known also as keyed hash functions, are a special type of hash functions which are used for message authentication and use keys.

An examples of existing MAC algorithms include the cipher-block chaining MAC (CBC-MAC) which work as follows: firstly, the sender divides the data $x$ into $n$-bit block $x_1, x_2, \ldots, x_t$. Let $E_k$ be the encryption using the algorithm $E$ and the key $k$, compute the block $H_t$ as follows: $H_1 \leftarrow E_k(x_1), H_i \leftarrow E_k(H_{i-1} \oplus x_i)$ where $2 \leq i \leq t$. General formulation of CBC-MAC is as follows: let $f$ be the underlying block cipher algorithm, $k$ be the shared key among parties, the message $x = x_1 x_2 \ldots x_t$ then the MAC is computed as $f_k^t(x) = f_k(f_k(\ldots f_k(f_k(x_1) \oplus x_2) \oplus x_{t-1}) \oplus x_t)$ [2]. To verify the MAC, the receiver computes the MAC again over the received message and admits the message if his computed MAC is equal to the received MAC from the sender. Otherwise, he reject the received message.

On the other hand, the aggregate MAC adds two functionalities: MAC aggregation and aggregate verification [23]. For $l$ different users with $l$ messages $(x^{(1)}, \ldots, x^{(l)})$, after generating the different MAC codes as $f_{k_1}^{x^{(1)}}, \ldots, f_{k_l}^{x^{(l)}}$, the aggregate MAC can be constructed by simply applying the XOR on the result MACs as follows: $f^{\text{agg}} = f_{k_1}^{x^{(1)}} \oplus f_{k_2}^{x^{(2)}} \oplus \cdots \oplus f_{k_l}^{x^{(l)}}$. The aggregate verification is performed as follows: first the receiver computes $f_{k_1}^{x^{(1)}} \oplus f_{k_2}^{x^{(2)}} \oplus \cdots \oplus f_{k_l}^{x^{(l)}}$ and compare it to the received $f^{\text{agg}}$. The receiver admits the messages if the result is equal to received one and reject if it is not equal.

Note that the difference between the aggregate MAC and aggregate signatures is that aggregate MAC works in the symmetric key encryption model while the signature works in the asymmetric key encryption model. This main difference lead to an overall difference in the operation and verification methods. Aggregate message authentication codes, according to Katz et al. [23], greatly reduce the communication overhead and very applicable to the ad-hoc networks. Since the LBS systems include many devices that are used in the ad-hoc network settings, we expect the aggregate MAC to be used widely in the LBS as well. The application scenario may include the fact that a single user would like to authenticate several users at once as the typical case of cluster formation required for anonymization. Typically, the MAC aggregation as a mean of message authentication based on symmetric model is more preferred due to the limited computational resources required for it over the public key cryptography which is computationally exhausting.

## 4.6. Homomorphic Encryption

The homomorphic encryption is a form of encryption where one can perform a specific algebraic operation on the plaintext by performing this algebraic operation on the ciphertext [43]. Homomorphic encryption techniques include modification of existing non-homomorphic schemes such like the unpadded RSA, El Gamal, Benaloh, etc. The most renown scheme is Paillier scheme. Details of how the homomorphic property is maintained including the following schemes.

- **Unpadded RSA [40]:** Let the public key be $m$ and $e$, then the encryption of message $x$ is given by $E(x) = x^e \mod m$ which satisfy the following homomorphic property $E(x_1)E(x_2) = x_1^e x_2^e \mod m = (x_1 x_2)^e \mod m = E(x_1 x_2 \mod m)$.
- **El Gamal [19]:** Let the public key be $p, g, h = g^a$, and $a$ is the secret key, then the encryption of a message $x$ is $E(x) = (g^r, x \cdot h^r)$ which satisfies the following homomorphic property: $E(x_1)E(x_2) = (g^{r_1}, x_1 \cdot h^{r_1})(g^{r_2}, x_2 \cdot h^{r_2}) = (g^{r_1+r_2}, (x_1 \cdot x_2)h^{r_1+r_2}) = E(x_1 x_2 \mod m)$.
- **Goldwasser Micali [21]:** Let the public key is the modulus $m$ and base $g$, and the encryption of message $E(x) = g^x r^m \mod m^2$, then we have the following homomorphic property: $E(b_1)E(b2) = r_1^2 x^{b_1} r_2^2 x^{b_2} = (r_1 r_2)^2 x^{b_1+b_2} = E(b_1 \oplus b_2)$ where $\oplus$ is the exclusive-or operation.
- **Paillier scheme [35]:** Let the public key be tge modulus $m$ and the base $g$, then the encryption of a message $x$ is $g^x u^r \mod m$ which satisfies the following homomorphic property: $E(x_1)E(x_2) = (g^{r_1} r_1^m)(g^{x_2} r_2^m) = g^{x_1+x_2}(r_1 r_2)^m = E(x_1 + x_2 \mod m)$

## 5. Other General Directions

As we mentioned before, the privacy in the LBS is still a challenging research direction that requires a lot of innovative solutions. In this section, we summarize a set of directions related to the aforementioned privacy architectures.

### 5.1. Membership Control

The membership control is an important directions for obtaining the trustworthiness in a distributed cooperative architecture aiming to provide privacy in LBS. Important elements of the membership control's research can be concentrated on the authentication, authorization and anonymization. Though these techniques are heavily studied and research in other settings, it need to be innovatively brought into the LBS system's field considering the different users' and applications' requirements and specifications.

## 5.2. Performance Consideration

The performance as an important issue not only in the LBS but in every system that seeks a commercial impact. However, in the LBS it is more critical to consider the performance since many platforms used for the LBS are basically mobile devices with limited resources. That is, related costs such like the computation, communication, and memory need to be considered for any successful design. Also, the privacy need to be considered as borderline when estimating the cost. The scalability as well is an important performance criterium that need to be considered.

## 5.3. Advancing the Secure Search

Heavy and intensive works have been performed on the search over encrypted data [1] which is a promising direction as the LBS requires a retrival for searchable data. However, since the LBS system mainly uses numerical data as an input for the location information, we expect the searchable encryption to be easier than general search encryption's case.

## 5.4. The Theory and Foundations

As a mater of fact, the definition of privacy is disputable [3]. Not only this but several security models guaranteeing privacy are, in many cases, not proved to provide privacy meeting the different definitions. Accordingly concrete privacy definitions are required. More precisely, research on the following directions would have fruitful results and impact:

1. Expressing privacy: allow user's awareness of both formally defined, and realistically expressible privacy.
2. Support various user's requirements: express privacy as a range of user requirements according to the operation modes reflexing need for the privacy at the time and location concerned.
3. More rules for the user: by allowing a "user defined" anonymization, grouping, etc, we can improve the adaptability of the LBS privacy.
4. Rigid definitions: not only for the defining the privacy but also for defining the privacy leakage or breach. That is, mechanisms are required for leakage quantification in LBS according to different real life scenarios.

## 6. Conclusion

Location based services are promising technological direction with several interesting research problems specially those related to the privacy as the LBSs are directly related to human information. To guarantee the privacy while providing a reasonable level of service in LBS, two approaches are used: cryptographic and non-cryptographic. Non-cryptographic approaches include cloaking, blurring, anonymization, among many others. On the other hand, the cryptographic approaches are not yet investigated though many cryptographic components studied in other networking settings are awaiting the deployment in the LBS systems. These cryptographic operations include mechanisms for secure multi party computation with extensions to the nature of the LBS's dynamic data and adversity, group formation (including authentication, signatures, group signatures, aggregate signatures and MAC, among others), and homomorphic encryption.

This paper introduced a general survey and many works are to be done. First, we will investigate the applicability of other SMC protocols and their variations for LBS devices (such like smartphones, PDAs, etc). Second, we will investigate the different scenario of applications and privacy breaches resulting from each in order to provide the proper cryptographic operations to limit or blocking these breaches.

## Acknowledgment

We we would like to thank Dr. Kuyung Chang and Dr. NamSu Jho for the discussion that motivated and helped writing this article. This work was supported by the IT R&D program of MKE/IITA. [2005-Y-001-04, Development of next generation security technology]